\documentclass[%
 reprint,
superscriptaddress,
nofootinbib,
 amsmath,amssymb,
prb,
longbibliography,
numerical,
]{revtex4-1}

\usepackage{graphicx}
\usepackage{dcolumn}
\usepackage{bm}
\usepackage{ulem}
\usepackage{color}
\usepackage{hyperref}

\newcommand{\bra}[1]{\left\langle #1 \right|}
\newcommand{\ket}[1]{\left| #1 \right\rangle}

\newcommand{\Hamiltonian}{{\cal H}} 
\newcommand{\id}{{\normalfont\hbox{1\kern-0.15em \vrule width .8pt depth-.5pt}}}

\newcommand{\pd}[2]{\frac{\partial #1}{\partial #2}}

\newcommand{\hide}[1]{}

\newcommand{\myspace}{\;\;\;\;\;\;\;\;\;\;\;\;\;\;\;\;\;\;\;\;\;\;\;\;\;}

\newcommand{\half}{\frac{1}{2}}
\newcommand{\ad}{a^\dagger}
\newcommand{\Kappa}{{\cal K}}

\begin{document}
\title{When do weak-coupling approaches accurately capture the dynamics of complex quantum systems?}
\author{Amir Fruchtman}
\affiliation{Department  of  Materials,  University  of  Oxford,  Oxford  OX1  3PH,  United  Kingdom}
\author{Neill Lambert}
\affiliation{CEMS, RIKEN, Saitama, 351-0198, Japan}
\author{Erik M. Gauger}\email[Corresponding author, ]{e.gauger@hw.ac.uk}
\affiliation{SUPA, Institute of Photonics and Quantum Sciences, Heriot-Watt University, EH14 4AS, United Kingdom}

\begin{abstract}
Understanding the dynamics of higher-dimensional quantum systems embedded in a complex environment remains a significant theoretical challenge. While several approaches yielding numerically converged solutions exist, these are computationally expensive and often provide only limited physical insight. Here we address the question when more intuitive and simpler to compute weak-coupling approaches still provide adequate accuracy. We develop a simple analytical criterion and verify its validity for the case of the much-studied FMO dynamics as well as the canonical spin-boson model.
\end{abstract}

\date{\today}

\pacs{Valid PACS appear here}
\maketitle

\section{Introduction}

Recent years have seen remarkable experimental progress in probing and controlling increasingly larger quantum systems in the condensed matter.\cite{Groblacher2015,Burek2014,Chan2011} In this setting it is often not possible to consider the environmental as only having a very small perturbative influence on the system of interest. Therefore, one cannot a priori expect conventional weak-coupling approaches~\cite{breuer2007theory}, attractive for their relative simplicity and interpretability, to remain a suitable tool for these systems.
However, master equations (MEs) based on a perturbative expansion nevertheless often provide good solutions in a variety of circumstances. \cite{} This includes, for example, comparison of such a model to experimentally observed excition-induced dephasing for laser-driven Rabi oscillation in semiconductor quantum dots \cite{Ramsay2010,Ramsay2010a}, or the close agreement of ME models with numerically exact solutions \cite{McCutcheon2011, Pollock2013, Shabani2012}. In some but not all of those cases, a judiciously chosen transformation allows a redefinition of system and environment before the perturbative expansion is performed (see, e.g. Refs.~\onlinecite{McCutcheon2011, Pollock2013}).

An important question to address then is: when is a weak-coupling approach `good enough' for capturing the essentials of the dynamics qualitatively or even on a quantitative level? Here, we develop a criterion for predicting when a weak-coupling is expected to perform well. 
Our criterion is based on a reasonably straightforward analytical expression that is, crucially, easy to evaluate, whilst also lending itself to an intuitive physical interpretation.

We consider two different weak coupling techniques: time convolutionless (TCL) master equations~\cite{breuer2007theory} and a second method based on the phase-space representation of the full density matrix\cite{Fruchtman2015} (P-mat). Interestingly, we show that both approaches give rise to exactly the same criterion, despite their rather different nature. This suggests that our criterion has applicability beyond a particular perturbative approach.

We apply both approaches to the canonical spin boson model\cite{Leggett1987} as well as the much studied FMO complex \cite{Adolphs2006,Ishizaki2009,Engel2007,Mohseni2008,Lambert2012,Nalbach2011b}. The latter has received a significant amount of attention and is a prime example of the complicated interplay between coherent dynamics interwoven with significant environmental influences. The advantage of this system is that a large body of literature and numerically converging methods exist. Interestingly, our criterion indicates that despite the relatively strong coupling, a weak-coupling treatment is appropriate at lower but not necessarily at higher temperatures. We note that the FMO problem has previously been tackled with weak-coupling techniques~\cite{Mohseni2008,Pachon2012,Ishizaki2010,Shabani2012,Shabani2014, Jeske2015}, but here we not only use novel method but also introduce a rigorous criterion for when such approaches are indeed permissible.

In the `grey area'  where weak-coupling is no longer strictly justified, we find that the quality of the different approaches differs. Some of us have previously found that the P-mat method outperformed the commonly used secular, second-order Born-Markov master equation\cite{Fruchtman2015}. For the examples studied here, we find that a TCL ME gives slightly better short time dynamics than P-mat, whilst the latter frequently performs better at longer times as the system approaches thermalisation. As expected, fourth-order TCL typically (but not always) beats second-order approaches, but may also lead to unphysical results in the strong-coupling regime.
To arrive at robust conclusions, we supplement comparisons of the population dynamics (as in Refs.~\onlinecite{Ritschel2011b, Shabani2012}) by a  trace distance, which also relies on agreement of the coherences.

The remainder of this paper is organised as follows: in Sec II we introduce the two aforementioned weak coupling approaches. We formally extend the P-mat technique from Ref.~\onlinecite{Fruchtman2015} to a multisite system with independent bosonic baths. Following Breuer \& Pettrucione\cite{breuer2007theory} we also briefly discuss the TCL technique. Section III presents the derivation of our weak coupling criterion. In Sec IV we describe HEOM, which is our benchmark technique for numerically converged solutions. Section V contains the key results of this work: the application of the criteria and plots of the trace distance, populations and coherences. We conclude with a summary of our results in Sec VI.

\section{Weak coupling techniques}
\subsection{Model}\label{subsection model}
As a starting point we use the `standard' open-systems Hamiltonian, which is given by
\begin{gather}\label{starting Hamiltonian}
\Hamiltonian = \Hamiltonian_S + \Hamiltonian_B + \Hamiltonian_I ~,
\end{gather}
where $\Hamiltonian_S$ is a finite-dimensional Hamiltonian for the system of interest, and $\Hamiltonian_B$, $\Hamiltonian_I$ are the bath and interaction Hamiltonians, respectively. The bath (or baths in the case of the FMO complex) is modelled by a collection of harmonic oscillators that are linearly coupled to the system according to:
\begin{align}
\Hamiltonian_B &= \sum_{\nu,k}\omega_{\nu,k} \ad_{\nu,k} a_{\nu,k} ~,\\
\Hamiltonian_I &= \eta\sum_\nu  V_\nu B_\nu = \eta \sum_{\nu,k} V_\nu g_{\nu,k} (\ad_{\nu,k}+a_{\nu,k}) ~.\label{linear interaction}
\end{align}
Here, $V_\nu$ are system and $B_\nu$ the bath operators with bath index $\nu$. The $g_{\nu,k}$ are the coupling constants, and $\ad_{\nu,k}$ the bosonic creation operators satisfying $[a_{\nu,k},a_{\nu',k'}] = [\ad_{\nu,k},\ad_{\nu',k'}] = 0$, $[a_{\nu,k},\ad_{\nu',k'}] = \delta_{k,k'} \delta_{\nu,\nu'}$.

Note that for the case of phonons, the justification for the linear form of (\ref{linear interaction}) 
usually hinges on a weak-coupling argument: We assume that the equilibrium position of the atoms hosting the vibrational modes only vary slightly when an electron or an exciton is present. Expanding the potential energy between these atoms to first order leads to the linear coupling term of Eq.~(\ref{linear interaction})\cite{Leggett1987, mahan2000many}. Conversely, for very strong coupling, one would not be able to justify the first-order approximation.

\subsection{The time convolutionless (TCL) technique} 
The TCL master equation is based on the projector-operator technique, stating that the system's dynamics obeys the time-local master equation 
\begin{gather}\label{master equation}
\pd{}{t}\rho_S(t) = \Kappa(t) \rho_S(t) ~,
\end{gather}
where the superoperator $\Kappa(t)$ is known as the TCL generator, and $\rho_S(t) = \text{Tr}_B(\rho)$ is the reduced density matrix of the system. 
It is time-local because $\rho_S$ on the righthand side of Eqn.~(\ref{master equation}) only features the current time $t$. All non-Markovian memory effects are thus contained within the TCL generator. A full derivation and discussion of it is found in chapter 9 of Ref.~\onlinecite{breuer2007theory}. 

Deriving an expression for the full TCL generator is as complex as solving the full von-Neumann equation for the system plus environment, so in practice we approximate it using a perturbative expansion in powers of the interaction\footnote{This is sometimes not possible, but should be possible for short times/weak coupling \cite{breuer2007theory}}:
\begin{gather}\label{kappa series}
\Kappa(t) = \sum_n \eta^n \Kappa_n(t).
\end{gather}
We use the Hamiltonian (\ref{starting Hamiltonian}), with a factorising initial condition $\rho(0) = \rho_S(0)\otimes\rho_B$, where $\rho_B = {\cal N}e^{-\beta \Hamiltonian_B }$, ${\cal N}$ is the environment partition sum and $\beta = 1/ (k_B T)$ inversely related to the temperature. For factorising initial condition and linear coupling, all odd terms in the TCL expansion vanish. The second and fourth terms are explicitly given by\cite{breuer2007theory,Jang2002}:
\begin{align}
\Kappa_2(t) = &-\sum_{\nu_0,\nu_1} \int_0^t dt_1 V_{\nu_0}^\times(t)R_{\nu_0,\nu_1}(t,t_1),\label{Kappa2 full}\\
\Kappa_4(t) = &\sum_{\nu_0,\nu_1,\nu_2,\nu_3} \int_0^t dt_1 \int_0^{t_1} dt_2 \int_0^{t_2} dt_3 \Big\{\nonumber
 \\&V_{\nu_0}^\times(t)\left[V_{\nu_1}^\times(t_1),R_{\nu_0,\nu_2}(t,t_2)\right]R_{\nu_1,\nu_3}(t_1,t_3)\notag\\
+&V_{\nu_0}^\times(t)\left[V_{\nu_1}^\times(t_1)R_{\nu_1,\nu_2}(t_1,t_2),R_{\nu_0,\nu_3}(t,t_3)\right]\Big\}.\label{kappa4 full}
\end{align}
Here
\begin{align}
V_\nu^\times(t) \square \equiv & [V_\nu(t),\square],\\
V_\nu^\circ(t) \square \equiv & \{V_\nu(t),\square \},\\
R_{\nu_a,\nu_b}(t_a,t_b)  \equiv & D^{\nu_a,\nu_b}(t_a-t_b)V_{\nu_b}^\times(t_b)\notag\\&+i D^{\nu_a,\nu_b}_1(t_a-t_b)V_{\nu_b}^{\circ}(t_b),\label{R superOperator}
\end{align}
and $D^{\nu_a,\nu_b}(t),D^{\nu_a,\nu_b}_1(t)$ are the real and imaginary part of this response function, respectively, given by 
\begin{align}\label{response matrix}
\alpha^{\nu_a,\nu_b}(t) &= D^{\nu_a,\nu_b}(t) + i D^{\nu_a,\nu_b}_1(t) \notag\\
& = \eta^2 \text{Tr}\Big\{B_{\nu_a}(t) B_{\nu_b}\rho_B \Big\}.
\end{align}

We note that in the strong-coupling regime, the procedure for calculating the TCL generator Eq.~(\ref{kappa series}) may fail\cite{breuer2007theory}.
Even when it does not, 
there is no guarantee that a truncation will still yield physical results. Indeed below we find strong-coupling examples where TCL4 [i.e.~truncating the series in Eq.~(\ref{kappa series}) after the fourth order term] results in a positive, and thus unphysical, eigenvalue of the generator in the long-time limit\footnote{We expect one steady-state with vanishing eigenvalue, and all the other eigenvalues should have a negative real part}.

\subsection{Multisite P-representation with independent baths}
The P-matrix approach~\cite{Fruchtman2015} is a different weak-coupling expansion for approximating the reduced system dynamics. In this case, one approximates the {\it generating function} of the time evolution generator as opposed to approximating the  {\it generator} like in the TCL technique. Interestingly, we shall find that this rather subtle distinction may have substantial impact even at second order.

In the P-matrix picture, we write down the dynamics of the reduced density matrix as
\begin{gather}
\rho(t) = U(t)e^{\Theta(t)}\rho(0),
\end{gather}
where $U(t)$ is the evolution operator for the dynamics of the closed system, and the effects of the environment and its memory are captured by the influence functional $\Theta(t)$. Following Ref.~\onlinecite{Fruchtman2015} we obtain a perturbative expansion of the influence functional in powers of the interaction parameter $\eta$, $\Theta(t) = \sum_n \eta^{n}\Theta_{n}(t)$.
We find that the influence functional expanded to second order bears a close formal relationship to the 
TCL expansion, and is simply given by
\begin{gather}\label{Pmat2}
\Theta_2(t) = \int_0^t d\tau \, \Kappa_2(\tau).
\end{gather}

Note that in Ref.~\onlinecite{Fruchtman2015} the P-matrix technique was developed for a single bath only, whereas Eq.~(\ref{Pmat2}) contains its extension to multiple independent baths.

\section{Weak coupling criterion}

The canonical definition of the coupling strength of an open system to its environment is the reorganization energy, i.e.~the potential energy associated with shifting the oscillator modes into their new equilibrium position in the presence of an excitation of the system,
\begin{gather} 
\lambda_\nu = \int_0^\infty d\omega \frac{J_\nu(\omega)}{\omega} ~, 
\end{gather}
where $J_\nu(\omega) = \eta^2 \sum_k g_{k,\nu}^2/\omega_{\nu,k} \delta(\omega-\omega_{\nu,k})$.
Note that the linear-coupling form of Eq.~(\ref{linear interaction}) implies the reorganisation energy should remain small (at least for the common case of a phonon bath).
Weak-coupling is sometimes defined by the condition $\lambda_\nu \ll V_{ij}$ where $V_{ij}$ represent off-diagonal coupling elements of the Hamiltonian between distinct basis states (typically chosen as the site basis)\cite{Ishizaki2009,Cheng2009}. Whilst straightforward for a two level system, in higher-dimensional systems such as the FMO complex, one may wonder exactly which $V_{ij}$ ought to be considered. Arguably, some off-diagonal terms in the Hamiltonian may be small or even vanish without automatically implying a strong coupling scenario. The authors of Ref.~\onlinecite{Ritschel2011b} write that in many cases the `` \ldots reorganization energy is not a reasonable  measure for the coupling strength \ldots'' because the system's dynamic frequencies are not taken into consideration. They suggest circumventing this shortcoming by defining an effective reorganisation energy $\lambda_\text{eff} = \int_{E_\text{min}}^{E_\text{max}} d\omega J(\omega)/\omega$, which only spans the relevant energy interval containing the system frequencies.

In this paper we analyse a different approach, which not only takes the system frequencies into account but also considers the temperature of the environment. 
Essentially, we let ourselves be guided by wishing to apply the term weak-coupling to situations when higher order expansion terms beyond the second order are not required for reliably capturing the open systems dynamics. To this end, we explicitly compare terms from a 4$^\text{th}$ order expansion to 2$^\text{nd}$ order terms, obtaining the following weak coupling criterion:
\begin{gather}\label{criterionFull}
\left| \frac{\bra{ii} \Kappa_4 \ket{ii}}{\bra{ii} \Kappa_2 \ket{ii}}\right|  \ll 1 ~,
\end{gather}
which must hold separately for each eigenenergy $i$, and where $\bra{ii} \Kappa_2 \ket{ii}$ and $\bra{ii} \Kappa_4 \ket{ii}$ are explicitly given in Eqs.~(\ref{K2approx}) and (\ref{eqkappa4ii}) of the Appendix. This ``full'' criterion, whilst evaluated straightforwardly enough, does not easily lend itself to providing much analytical insight. Therefore, we also consider a ``simplified'' version that is more amenable to physical interpretation. This simplified criterion reads 
\begin{gather}
\label{criteria}
\sum_{j\neq i} \Upsilon_{ij} \ll 1 ,~
\end{gather}
where 
\begin{gather}
\label{upsilons}
\Upsilon_{ij} = 2 |V_{ij}|^2 \int_0^t \tau d\tau D(\tau) \cos\Delta_{ij}\tau ~.
\end{gather}
Here, $|V_{ij}|^2 = \sum_\nu \left|\left\langle i |V_\nu|j \right\rangle\right|^2$ with $\ket{j}$ being the eigenstates of the system, $\Delta_{ij} = \epsilon_i-\epsilon_j$ is the energy difference between two eigenenergies $i,j$, and we have assumed that all spectral densities take the same form $J_\nu(\omega) = J(\omega)$.\footnote{Generalising for different spectral densities is straightforward.} For a thermal environment we have $D(t) = 
\int_0^\infty d\omega J(\omega)\coth(\beta \omega / 2) \cos(\omega t)$. Further, $t$ is the timescale of interest (i.e.~the duration for which we want the calculations to remain accurate). When $D(\tau>t)\approx 0$, it is justified to take the upper limit of the integral to infinity. We provide the explicit derivation of Eq.~(\ref{criteria}) in the Appendix, and note that TCL and P-mat both lead to exactly the same final expression.

For the examples studied in the following sections of this paper, the simplified criterion of Eq.~(\ref{criteria}) turns out to be similarly stringent as the full criterion. However, not having been able to mathematically prove that it will always be sufficiently rigorous, we suggest using it carefully and supplementing it by Eq.~(\ref{criterionFull}) if in doubt.
Eq.~(\ref{upsilons}) is closely related to the known phenomenon of slippage of initial conditions\cite{Suarez1992,Gaspard1999}, and here we argue that this slippage is directly related to the coupling strength, or more precisely to the validity of the perturbative expansion. Note that appreciable slippage of initial conditions marks the onset of non-Markovian effects, rendering a perturbative second-order expansion (and thus our definition of weak coupling) invalid.

We shall now briefly introduce a technique yielding numerically converged results, which will in the following serve as benchmark against which we may compare our perturbative expansions, and provide evidence for the validity and usefulness of the above criteria.

\section{HEOM}

The hierarchical equations of motion (HEOM), first proposed by Tanimura and Kubo\cite{Tanimura3, Tanimura2, Tanimura}, map the exact equation of motion of the reduced system density matrix to a simpler set of equations describing a series of coupled auxiliary density matrices.  This series is arranged in a hierarchy proven to converge. Derivation of this hierarchy requires that the response function of environment can be described by a sum of exponential terms.  Examples of spectral densities where this is analytically the case include the Drude-Lorentz spectral density\cite{Tanimura3}, a Lorentzian spectral density \cite{Ma2012} and an underdamped Brownian oscillator \cite{tan1,Tan2,Tan3,arend}.

In the FMO example below we use a Drude-Lorentz spectral density. However, in general one can also numerically fit the response function of an arbitrary bath to a sum of exponential terms, similar in spirit to Ref.~\onlinecite{Dattani2012}. We find that this approach works well for both a Ohmic and super-Ohmic spectral densities.
Due to the form of the HEOM, it is convenient to fit the real and imaginary parts of the response function independently, defining in general $\alpha(t) = D(t) + i D_1(t)$, with
\begin{eqnarray}
D(t) &=& \sum_{k=1}^{N_R} c^R_k e^{-\mu^R_k t},\\
D_1(t) &=& \sum_{k=1}^{N_I} c^I_k e^{-\mu^I_k t}.
\end{eqnarray}
Then, when constructing the HEOM for such general spectral densities, we distinguish between those ancilliary density operators originating from the real and imaginary parts of the response function, and write the hierarchy index $\vec{n}=(\vec{n}_{\nu R},\vec{ n}_{\nu I})$. Here $\vec{n}_{\nu j}=(n_{\nu j1},n_{\nu j2},..,n_{\nu jN_j})$, $n_{\nu jk}$ are integers $n_{\nu jk} \in\{0,..,N_c\}$, up to the cut-off tier of the hierarchy  $N_c$, and where $\nu  \in\{0,..,N\}$ labels the different independent baths, up to their number $N$. For notational simplicity we assume all baths have the same correlation functions, but again generalisation is straight-forward. The HEOM (with renormalized coupling between hierarchies\cite{shi09}) then takes the form

\begin{eqnarray}
\dot{\rho}^n(t)&=&-\left(i H_S^{\times} + \sum_{\nu=1}^N\sum_{j=R,I}\sum_{k=1}^{N_j} n_{\nu jk}\mu^j_k \right) \rho^n(t) \\
&&- i \sum_{\nu=1}^N\sum_{k=1}^{N_R} \sqrt{\frac{n_{\alpha Rk}}{|c^R_k|}}c^R_k V^{\times}_{\nu }\rho^{n_{\nu Rk}^-}(t)\notag\\
&&+ \sum_{\nu=1}^N\sum_{k=1}^{N_I} \sqrt{\frac{n_{\nu Ik}}{|c^I_k|}}c^I_k V^{\circ}_{\nu }\rho^{n_{\nu Ik}^-}(t)\notag\\
&&-i \sum_{\nu=1}^N\sum_{j=R,I}\sum_{k=1}^{N_j} \sqrt{(n_{\nu jk}+1)|c^j_k|} V^{\times}_{\nu } \rho^{n_{\nu jk}^+}(t).\notag
\end{eqnarray}

The notation $\rho^{n_{\nu jk}^{\pm}}$ indicates an increase/decrease of just the $\nu jk$'th element of the hierarchy index by $1$.  One then solves these coupled equations numerically, setting all the ancillary ($\vec{n}\neq (0,0,...,0)$) density matrices to zero at $t=0$. 

\section{Results}

\subsection{Spin-boson model example}

\label{Appendix 2LS}
\begin{figure}
\centering
\includegraphics[scale=0.75]{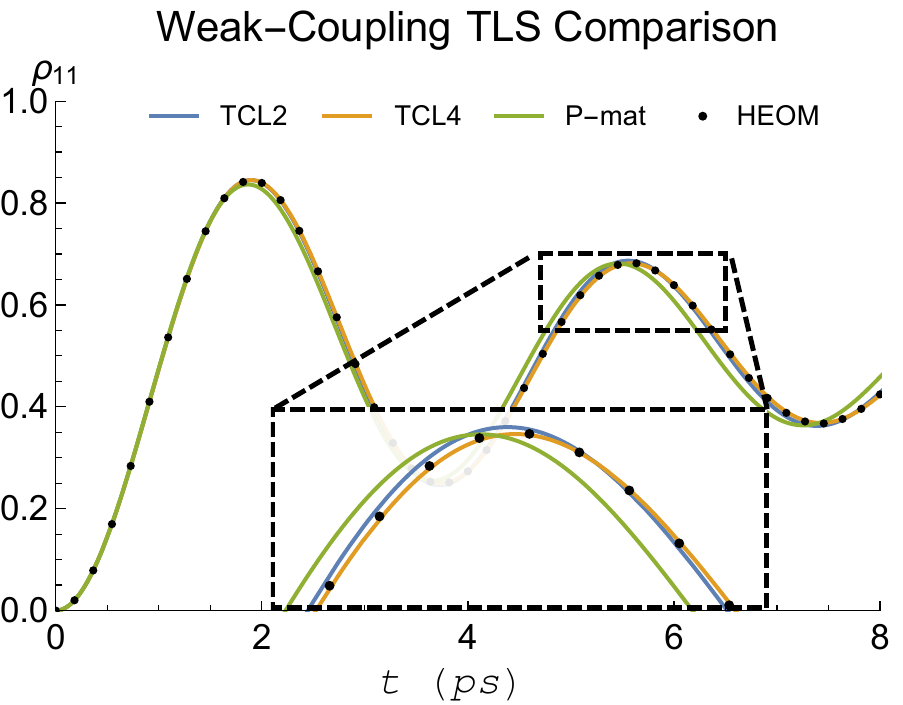}
\caption{\label{figure TLS compare}
Comparison of the spin boson dynamics calculated using HEOM (dashed) to our perturbative techniques (solid). Here we consider the weak-coupling case with $\eta = 1$ (see main text for other parameters).
}
\end{figure}

\begin{figure}
\centering
\includegraphics[scale=0.75]{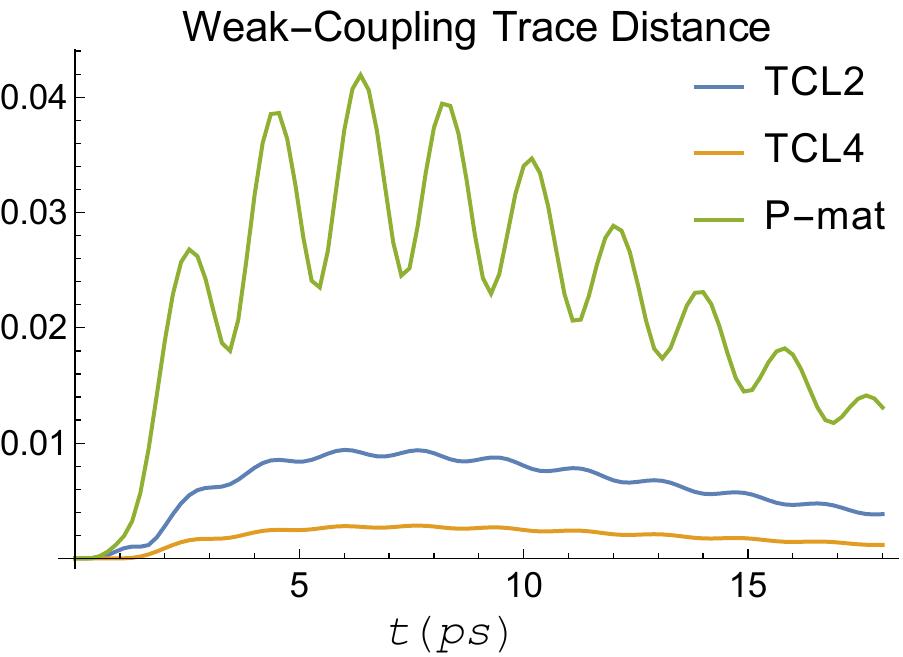}
\caption{\label{figure TLS traceDistance}
Trace distance between the spin boson dynamics calculated using HEOM and the different perturbative techniques. Again, we consider a clearly cut weak coupling scenario with parameters as in Fig.~\ref{figure TLS compare}.
}
\end{figure}

\begin{figure}
\centering
\includegraphics[scale=0.75]{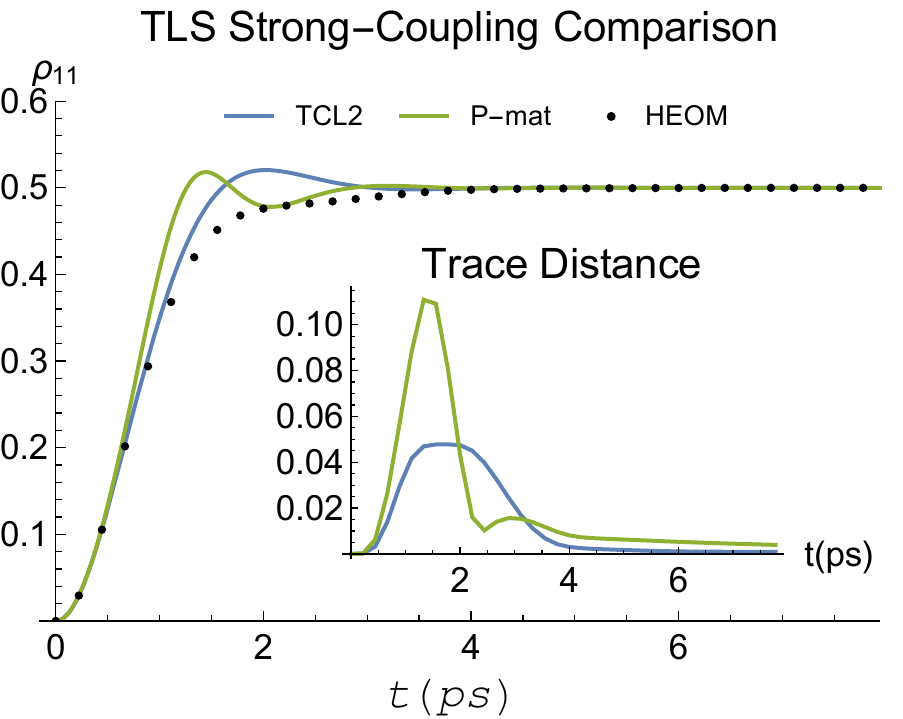}
\caption{\label{figure TLS strong}
Comparison between the spin boson dynamics calculated using HEOM and weak-coupling techniques, for a strong-coupling case, $\eta = 10$. Other parameters as in Fig.~\ref{figure TLS compare}.
}
\end{figure}

As a first case study, we apply our criteria to the canonical example of the spin-boson model\cite{Leggett1987} with Hamiltonian given by:
\begin{gather}
\Hamiltonian = \half \Delta \sigma_x + \sum_k \omega_k \ad_k a_k + \sigma_z \sum_k g_k (\ad_k+a_k) ~,
\end{gather}
where $\sigma_x,\sigma_z$ are the usual Pauli matrices and $\Delta$ represents the Rabi frequency of the spin. Further, $a^{(\dagger)}$ denotes the annihilation (creation) operator of a bosonic mode $k$ with frequency $\omega_k$, and $g_k$ are the spin-boson coupling elements. We consider an Ohmic spectral density with exponential cutoff as follows
\begin{gather}
J(\omega) = \sum_k g_k^2 \delta(\omega-\omega_k) = \eta \lambda (\omega/\omega_c) e^{-\omega^2/\omega_c^2} ~,
\end{gather}
where $\eta$ is a dimensionless parameter allowing us to easiliy interpolate between weak and strong coupling.
We choose $\Delta = \pi/2$ ps$^{-1}$, $T = 50$ K, $\lambda = 0.01485$ ps$^{-1}$, and $\omega_c = 2.2$~ps$^{-1}$. 

Evaluating our weak-coupling criteria Eqs.~(\ref{criteria}, \ref{criterionFull}) for these parameters gives 0.04 $\eta$ (0.06 $\eta$) for the simplified (full) version of the criterion, meaning the system is well into the weak-coupling regime when we set $\eta=1$. This can be seen in Figs.~\ref{figure TLS compare} and \ref{figure TLS traceDistance} which show a comparison of TCL2, TCL4, and P-mat techniques with the numerically exact HEOM dynamics. As expected the perturbative techniques capture the dynamics very accurately in this case. P-mat performs worst in this case (despite having previously been shown to outperform a standard secular Born-Markov master equation in a comparable scenario\cite{Fruchtman2015}), although the gap is narrows towards longer times when the system approaches thermal equilibrium. The trace distance of all three perturbative approaches features oscillations at twice the Rabi frequency of the natural precession time of the spin, clearly visible in Fig.~\ref{figure TLS traceDistance}. 
The trace distance between two density matrices is defined by
\begin{gather}
\mathcal{T}(\rho_1,\rho_2) = \frac{1}{2}Tr\{\sqrt{(\rho_1-\rho_2)^2}\} = \frac{1}{2}\sum_i |\lambda_{i}|,
\end{gather}
where $\lambda_i$ is the i'th eigenvalue of $(\rho_1-\rho_2)$.

When we increase the system and environment coupling by setting $\eta=10$, our criteria indicate that we can no longer expect to be in the weak-coupling regime. In Fig.~\ref{figure TLS strong} we show a comparison between TCL2 and P-mat techniques with exact HEOM calculations, and it is apparent that both fail not only quantitatively but also qualitatively at intermediate times around $t \approx 1.5$~ps. We do not show TCL4 in this case because the TCL4 generator features an unphysical positive eigenvalue in the long time limit.

\subsection{FMO complex dynamics}

We now turn to the dynamics of the Fenna-Matthews-Olsen (FMO) complex -- a prime example of complex quantum dynamics in the tricky regime between weak and  strong environmental coupling in a higher-dimensional Hilbert space. We follow the 7-site FMO model considered by Ishizaki and Fleming~\cite{Ishizaki2009}, where all chlorophylls have the same environment given by a Lorentzian spectral density $J_\nu(\omega) = \frac{2}{\pi} \lambda \frac{\gamma \omega}{\omega^2+\gamma^2}$ with $\lambda = 6.59$ ps$^{-1}$. We consider the three cases for the environmental parameters also discussed in Ref.~\onlinecite{Ishizaki2009}, namely $T=77~\mathrm{K}, \gamma^{-1} = 50~\mathrm{ps}; T=300~\mathrm{K}, \gamma^{-1} = 50~\mathrm{ps}$; and $T=300~\mathrm{K}, \gamma^{-1} = 166~\mathrm{fs}$. Once again comparing exact HEOM results to perturbative expansions, we will find that our criteria predict the validity of the expansions not only for the spin boson model but also for this much more involved case.

Table \ref{table criterion} lists of the values of the criteria for the different sets parameters mentioned above. Here we observe a factor of roughly two between the simplified, Eq.~(\ref{criteria}) and full criterion, Eq.~(\ref{criterionFull}), with the simplified version being the more stringent one. However, we note that both agree in their  classifications of weak vs strong coupling scenarios. Further, the reorganisation energy is constant across the three cases as it does not depend on the temperature or the cut-off frequency, and is thus clearly not a good measure for the coupling strength.
\begin{table}
\begin{center} 
    \begin{tabular}{ | l | c | c | c | c |}
    \hline
    Temperature & $\gamma^{-1}$ & Max Full & Max Simplif- & $ \lambda_n /$ \\ 
	     &	&	Criterion	& ied Criterion	& max $|V_{12}|$ \\    
    \hline
    77~K & 50~fs & 0.04 & 0.09 & 0.4
    \\ \hline
    300~K & 50~fs & 0.19 & 0.38 & 0.4 
     \\ \hline
    300~K & 166~fs & 1.09 & 2.6 & 0.4
     \\ \hline
    \end{tabular}
\end{center}
\caption{\label{criterion table}
Full and simplifed criteria applied to different FMO configurations. In each case we choose the largest value of LHS of Eqns.~(\ref{criterionFull}) and (\ref{criteria}) over all $i$. The reorganisation energy, $\lambda_n = 6.59$~ps$^{-1}$, does not depend on the cut-off frequency $\gamma$ or the temperature, and we show the value of the reorganisation energy divided by the largest dipole-dipole coupling in the system, $|V_{12}| = 16.5$~ps$^{-1}$ (note that here 1 and 2 refer to the site basis, whereas throughout the rest of this paper we use the energy eigenbasis).
}\label{table criterion}
\end{table}

\begin{figure}
\centering
\includegraphics[scale=0.8]{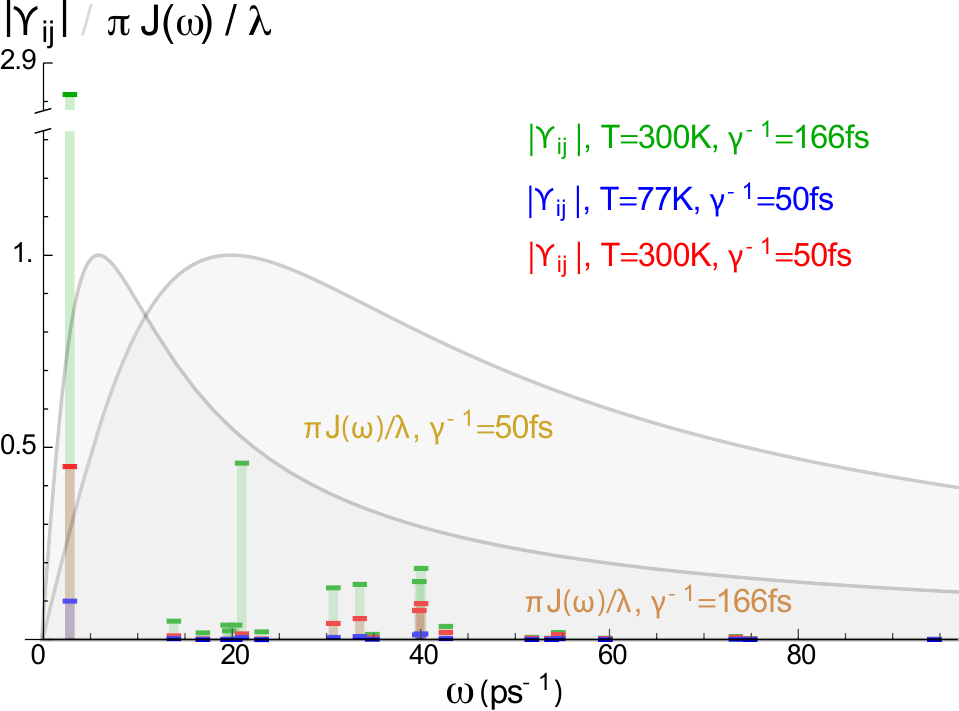}
\caption{\label{figure Strength Visualise}
Visualisation of the simplified criterion: the different $|\Upsilon_{ij} |$ are shown at their corresponding frequencies $\Delta_{ij}$. The rescaled and dimensionless respective spectral densities, $\pi J(\omega)/\omega$, are also shown as the coloured background areas, to illustrate at which frequencies environmental effects are expected to be dominant.
}
\end{figure}

We visualise the $\Upsilon_{ij}$ of Eq.~(\ref{criteria}) for all three FMO cases in Fig.~(\ref{figure Strength Visualise}). Notably, two of the system eigenstates  are almost resonant, being split by only $\sim 2.8~\mathrm{ps}^{-1}$. This pair experiences strong environmental interactions for the $\gamma^{-1} = 166$~fs cutoff ($\Upsilon_{ij}\sim 2.8$), clearly placing the system into the strong-coupling domain according to our definition. By contrast,  for $\gamma^{-1} = 50~\mathrm{fs}$ this pair sits in an intermediate regime (at $T=300$~K), whereas all other occurring frequencies satisfy our weak-coupling criterion.
Consequently, we would expect a very good agreement between exact numerics and weak-coupling techniques in the low-temperature cases at $T=77$~K, but at $T=300$~K we expect the weak-coupling techniques to work only for $\gamma^{-1}=50~$fs and not for $\gamma^{-1}=166$~fs. We shall see that these expectations are met by our dynamical simulations discussed in the following.

To perform our simulations, the real part of the response function $D(t)$ is approximated by a single exponent and a delta function\footnote{The imaginary part of the response function has a single exponent as well}
\begin{align}
D(t)  \rightarrow 2 & \lambda k_b T \left(1-\frac{2\gamma^2}{(2\pi k_b T)^2-\gamma^2}\right)e^{-\gamma t} \\&+\frac{ 8\lambda k_b T \gamma }{(2\pi k_b T)^2-\gamma^2}\delta(t).
\end{align}
This approximation is made to overcome memory restrictions of the HEOM implantation. For a fair comparison, we use the same response function for the TCL and P-mat methods, though these methods are not restricted to a certain structure of the response function. Attempting to capture the full density matrix dynamics as opposed to just the evolution of the populations, we shall also plot and discuss the trace-distance between the perturbative solutions and the HEOM benchmark. 

\begin{figure*}
\centering
\includegraphics[width=\textwidth]{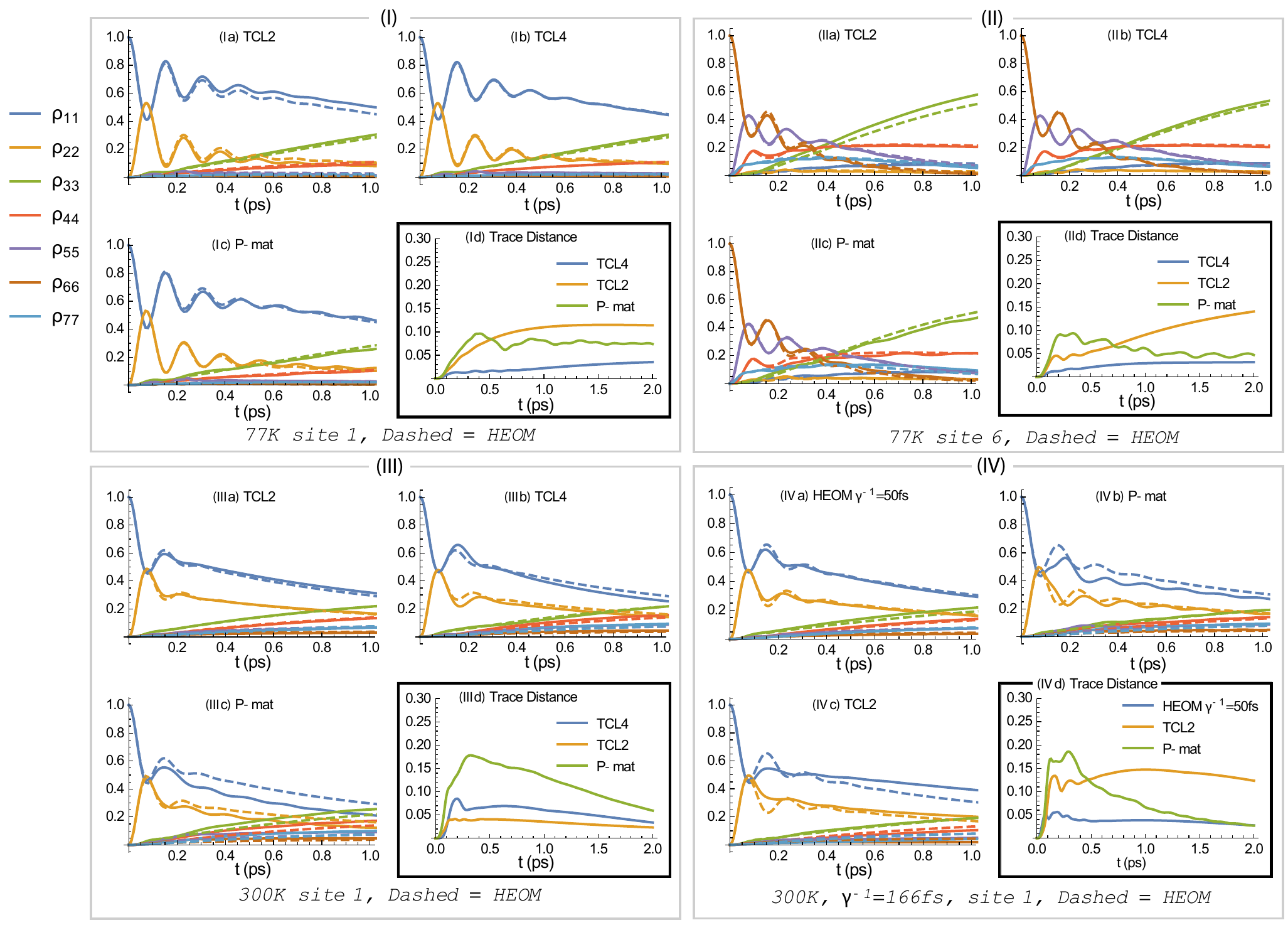}
\caption{\label{figure FMO compare} Benchmarking of dynamics obtained from perturbative techniques (solid) against numerically converged HEOM (dashed). Panels (I-III): Initial state and temperature as indicated at the bottom of each panel. Subplots (a)-(c) show comparisons of TCL2, TCL4, and P-mat, respectively, vs HEOM. Subplots (d) give the trace distance between HEOM and the aforementioned perturbative approaches (extending out to longer times). All other parameters are as in Ref.~\onlinecite{Ishizaki2009} with cutoff frequency $\gamma^{-1} = 50$~fs. 
(IV) FMO dynamics comparison with a stronger coupling: Subplot (a) contrasts HEOM for cutoff frequency $\gamma^{-1} = 166$~fs against HEOM for $\gamma^{-1} = 50$~fs. Subplots (b) and (c) show P-mat and TCL2 (solid) against HEOM, all using cutoff frequency $\gamma^{-1} = 166$~fs. Notably, P-mat performs worse than TCL2 at short times but then improves. TCL4 provides unphysical solutions and has therefore not been included.
}
\end{figure*}

Panels I-III of Fig.~\ref{figure FMO compare} show the trace distances and  populations from the weak-coupling techniques against the numerically exact HEOM calculations, all using $\gamma^{-1}=50$~fs and for $T=77 ,300$~K as indicated. We find that TCL2 compares favourably to P-mat at short times, while for longer times, with the notable exception of Fig.~\ref{figure FMO compare} III, P-mat becomes more accurate. We suspect that this is because in the P-mat formalism, the secular approximation is inherently included. By contrast, in TCL a full or partial secular approximation\cite{Jeske2015} still needs to be performed explicitly to guarantee that the system will go to the thermal state with respect to the temperature of the bath\footnote{provided there is no decoherence-free subspace\cite{Aharony2012} and only a single temperature.}. Note, however, that in a stronger-coupling regime the system is no longer expected to fully evolve into its thermal state\cite{Iles-smith2014}.

We conclude that in all these three cases, our weak-coupling techniques capture the relevant oscillations in the dynamics well, as expected. Rather surprisingly, in Fig.~\ref{figure FMO compare}III ($\gamma^{-1} = 50$~fs and $T = 300$~K) when using site 1 as the initial state, TCL2 outperforms not only P-mat but also TCL4. We note that this is not the case in the same configuration but with site 6 as the location of the initial excitation.

We now consider the lower cutoff frequency $\gamma^{-1} = 166$~fs at  $T=300$~K, having already identified this case as one which violates our weak-coupling criterion. As shown in Fig.~\ref{figure FMO compare}IV,  exact numerics suggest that the difference in the dynamics is deceptively small between $\gamma^{-1} = 50$~fs and 166~fs,  borne out both by the populations [subplot (a)] and the trace distance [subplot (d)]. Nonetheless, according to Table \ref{criterion table}, the two cases are vastly different from the point of view of the perturbation series, and indeed we find that TCL2 is overestimates the damping and does not capture the oscillations, whilst  P-mat underestimates the damping and shows more oscillations than HEOM. We do not present TCL4 results in this plot because in this regime, the TCL4 generator once more possesses an unphysical eigenvalue in the long-time limit. 

\section{Conclusion}

We have discussed a straightforward criterion for the often difficult decision when weak coupling approaches perform adequately for capturing higher-dimensional quantum dynamics in complex environments. We have discussed a rigorous as well as a simplified version of the criterion, with the latter constituting essentially a measure of the degree of `slippage of initial conditions' \cite{Suarez1992,Gaspard1999}. 

By presenting a comparative numerical study of two different classes of weak-coupling methods, contrasted against numerically converged HEOM results, we have verified the validity and predictive power of both variants of the criterion. We have used the canonical spin boson model and the much-studied energy excitation dynamics in the FMO complex as two representative examples to make this case.

Interestingly, we have identified two room temperature FMO configurations with almost identical HEOM results, whilst our perturbative solutions diverge significantly. This discrepancy is captured by our criterion which confirms that despite the apparent similarity of numerically converged results, one configuration sits well in the weak-coupling regime, whilst the other is sufficiently strongly coupled such that a perturbative approach is no longer appropriate.

Finally, we have here extended the P-mat approach, first introduced in Ref.~\onlinecite{Fruchtman2015}, to a situation with multiple baths, and discussed its relationship to time-convolutionless master equations. We have seen that P-mat stands its ground reasonably well compared to TCL2, particularly in the long-time limit. Whilst not the subject of this study, we note that the concept of tiered environments\cite{Fruchtman2015} can also be straightforwardly applied to higher-dimensional systems and multiple baths.

\begin{acknowledgements}
We thank Simon Benjamin, Brendon Lovett, and Franco Nori for stimulating discussions. 
This work was supported by the Leverhulme Trust (RPG-080). 
AF thanks the Anglo-Israeli association and the Anglo-Jewish association for funding.
NL is partially supported by the FY2015 Incentive Research Project.
EMG acknowledges funding from the Royal Society of Edinburgh and the Scottish Government.
\end{acknowledgements}

\bibliography{mybiblio,nlbib}

\appendix
\section{Derivation of the criterion}
\label{section derivation}
In this section we derive the criterion (\ref{criteria}) in the main text. For this analysis we look at the case where all of the different baths are independent, as defined in \ref{subsection model}, and they are all coupled to the system in the same manner, i.e. $J_{\nu}(\omega) = J(\omega)$ for all $\nu$. This simplifies Eqs.~(\ref{Kappa2 full}-\ref{kappa4 full}) with $R_{n_a,n_b}(t_a,t_b) = \delta_{n_a,n_b}R_{n_a,n_a}(t_a,t_b)$, and $D^{n_a,n_b}(\tau) = \delta_{n_a,n_b}D(\tau), D_1^{n_a,n_b}(\tau) = \delta_{n_a,n_b}D_1(\tau), \alpha^{n_a,n_b}(\tau) = \delta_{n_a,n_b}\alpha(\tau)$.

Working in the Liouville Space $\Hamiltonian_S \otimes \Hamiltonian_S$, with $\ket{i}\bra{j}\rightarrow \ket{ij}$, where $\ket{i}$ are the system's energy eigenstates, it is straightforward to write down an explicit expression for $\Kappa_2$:
\begin{align}
\bra{i j} &\Kappa_2 \ket{rs} = -e^{-i(\Delta_{rs}-\Delta_{ij})t}\sum_\nu\int_0^t d\tau\Big\{
\notag\\ & \sum_k \Big[\delta_{js} e^{-i\Delta_{k r}\tau}V^\nu_{ik}V^\nu_{kr}\alpha(\tau) + \delta_{ir} e^{+i\Delta_{k s}\tau}V^\nu_{sk}V^\nu_{kj}\alpha^*(\tau) \Big]
\notag\\&-V^\nu_{ir}V^\nu_{sj}\Big[e^{-i\Delta_{ir}\tau}\alpha(\tau)+e^{+i\Delta_{js}\tau}\alpha^*(\tau)\Big]\Big\},
\end{align}
where $\Delta_{ij} = \epsilon_i-\epsilon_j$, and $V^\nu_{ij} = \bra{i}V_\nu\ket{j}$.

The dephasing, decoherence, and Lamb-shift rates are given by matrix elements in the superoperator $\Kappa(t)$ that do not oscillate with $t$ as $t\rightarrow\infty$, meaning that in $\Kappa_2$ only elements with $\Delta_{rs}-\Delta_{ij} = 0$ contribute to the rates. In the same manner one can show that the $t$ dependence of the fourth order is $\bra{ij}\Kappa_4\ket{rs} \sim e^{-i(\Delta_{rs}-\Delta_{ij})t}$ for times much longer than the bath's memory time.
Assuming that the energies are non-degenerate in the broad sense, meaning $\sum_i(\epsilon_{n_i}-\epsilon_{m_i}) = 0 \leftrightarrow \{n_i\}=\{m_i\}$, then the only matrix elements that have these rates are either diagonal terms
\begin{align}\label{2nd order}
\bra{ij}\Kappa_2 &\ket{ij} = -\sum_\nu \int_0^t d\tau \Big\{  (V^\nu_{ii}-V^\nu_{jj})[V^\nu_{ii}\alpha(\tau)-V^\nu_{jj}\alpha^*(\tau)] \notag
\\
&+\sum_{k\neq i}|V^\nu_{ki}|^2 e^{-i\Delta_{ki}\tau}\alpha(\tau)
+\sum_{k\neq j}|V^\nu_{jk}|^2 e^{-i\Delta_{jk}\tau}\alpha^*(\tau)
\Big\}~,
\end{align}
or elements between two different projectors,
\begin{align}
\bra{ii}\Kappa_2\ket{jj}_{i\neq j} =& 
\\ \notag
\sum_\nu\int_0^t d\tau & |V^\nu_{ij}|^2\left[e^{-i\Delta_{ij}\tau}\alpha(\tau)+e^{+i\Delta_{ij}\tau}\alpha^*(\tau)\right]~.
\end{align}

The part proportional to the diagonal of the coupling operator $V$ is known in the literature as ``pure dephasing''. Incidentally because our model only considers linear coupling between the system and environment, if there were only diagonal terms in the interaction, then the exact contributions of these terms are equal to their second order expansion~\cite{Skinner1986}.

 Moreover, we are only interested in the real parts of the rates, as the imaginary parts are responsible for the Lamb shift. Finally, in Eq.~(\ref{2nd order}), the rates for $i\neq j$ are the ones for which the off-diagonal elements of the density matrix decay. For the sake of this paper we focus on the dephasing rates, i.e.~the rates in which the diagonal elements of the density matrix decay, i.e. 
\begin{gather}
\bra{ii} \Kappa_2 \ket{ii} =
\\\notag -2\sum_n \int_0^t d\tau \sum_{k\neq i} |V^n_{i,k}|^2\left[D(\tau)\cos(\Delta_{ik}\tau)-D_1(\tau)\sin(\Delta_{ik}\tau)\right]~.
\end{gather}
The part proportional to $D_1 = \text{Im}(\alpha)$ in the above equations is normally very small and could be ignored for the purpose of our discussion, hence we approximate the second order rates as
\begin{gather}\label{K2approx}
\bra{ii} \Kappa_2 \ket{ii} \approx
 -2\sum_\nu \int_0^t d\tau \sum_{k\neq i} |V^\nu_{i,k}|^2 D(\tau)\cos(\Delta_{ik}\tau)~.
\end{gather}
Defining
\begin{gather}\label{gamma def}
\Gamma_{t}(i,j) = \sum_{\nu}|V^\nu_{i,j}|^2\int_0^t d\tau D(\tau) \cos(\Delta_{ij}\tau)~,
\end{gather}
we obtain the main quantity of interest when comparing the second order and the fourth order of the rates.
In the Markovian limit, where the time of the experiment is much longer than the memory-kernel decay time $t \gg t_c$, we take the integral in Eqn.~(\ref{gamma def}) to infinity and get the usual rates
\begin{gather}
\Gamma_{\infty}(i,j) = \sum_{\nu}|V^\nu_{i,j}|^2 \frac{\pi}{2} J(|\Delta_{ij}|) \coth(\beta |\Delta_{ij}|/2)~.
\end{gather}
For times comparable or shorter than the decay time of the response kernel $D(t)$, or for the case where the spectral density is not smooth and has very sharp peaks, it is not justified to take the Markovian limit and one should evaluate the rates (\ref{gamma def}) for the time of the experiment.

Using a similar analysis on the fourth-order rates, we find that the correction to the rates introduced by a fourth order treatement are given by
\begin{align}
\bra{ii} \Kappa_4 & \ket{ii} \approx \myspace\myspace \label{eqkappa4ii}
\\\notag
-2 \sum_{\nu_1,\nu_2} & \int_0^t d\tau_1 \left[\int_0^{t-\tau_1} \tau_2 d\tau_2 +\int_{t-\tau_1}^t (t-\tau_2) d\tau_2\right] \times
\\ \notag & D(\tau_1) D(\tau_2)\times 
\\ \notag
&\sum_{k\neq i, p\neq i}|V^{\nu_1}_{ik}|^2 |V^{\nu_2}_{ip}|^2  \cos(\Delta_{ik}\tau_1+\Delta_{ip}\tau_2)(1+\delta_{kp})
\\ \notag
+2 \sum_{\nu_1,\nu_2} & \int_0^t d\tau_1 \int_0^{\tau_1} d\tau_2 (\tau_1-\tau_2)D(\tau_1) D(\tau_2)\times  
\\\notag
\Big[\sum_{i\neq k} & |V^{\nu_1}_{ik}|^2 V_{ii}^{\nu_2}(V_{ii}^{\nu_2}-V_{kk}^{\nu_2})\cos(\Delta_{ik}\tau_1) +
\\\notag +\sum_{k\neq i} & \sum_{p\neq\{i,k\}}|V^{\nu_1}_{ik}|^2|V^{\nu_2}_{kp}|^2 \cos(\Delta_{ik}\tau_1+\Delta_{kp}\tau_2)\Big] ~.
\end{align}
In the above equation we only kept terms that have $\tau_1$ or $\tau_2$ in their integrand. We note that there are other terms, namely terms with integrands similar to $(e^{-i \Delta \tau} - 1)/\Delta$ where $\Delta$ is an eigenfrequency of the system, which are generally, but not always, smaller than their $\tau$ equivalents derived by taking the limit $\Delta \rightarrow 0$.
Further, we ignore contributions by $D_1$ as in Eq.~(\ref{K2approx}).
Also, we assume that the system's energies and frequencies are not degenerate in the sense that $\sum_i E_i = \sum_j E_j \Rightarrow \{E_i\} = \{E_j\}$.

We note that Eq.~(\ref{eqkappa4ii}) consists of two terms. We conjecture (showing that this is true for the case of the spin-boson model and the FMO complex) that both terms are of the same order. For the sake of this analysis we can therefore ignore the second term. 

Now, because we are looking for a criteria for when the fourth order contribution is small, we can make it larger and take $1+\delta_{kp} \rightarrow 2$, and take the upper limit of the $\tau_2$ integral to $t$. Further, we expand the cosine and ignore terms proportional to $\tau_2 \sin(\Delta_{ik}\tau_1)\sin(\Delta_{ip}\tau_2)$ , which are smaller than their cosine equivalents, and we are left with
\begin{align}
\bra{ii} \Kappa_4 \ket{ii} \sim &  \\\notag
-4 \sum_{\nu_1,\nu_2} \int_0^t &d\tau_1 \int_0^{t} d\tau_2  \tau_2 D(\tau_1) D(\tau_2) \\\notag\sum_{k\neq i, p\neq i}&|V^{\nu_1}_{ik}|^2 |V^{\nu_2}_{ip}|^2\cos(\Delta_{ik}\tau_1) \cos(\Delta_{ip}\tau_2) ~.
\notag
\end{align}
This leaves us with the following criteria for when weak-coupling is a good approximation:
\begin{gather}\label{criteria calc}
\frac{\bra{ii} \Kappa_4 \ket{ii}}{\bra{ii} \Kappa_2 \ket{ii}} \lesssim 2 \sum_\nu \sum_{k\neq i} |V_{ik}^\nu|^2 \int_0^t d\tau \tau D(\tau) \cos(\Delta_{ik} \tau) \ll 1 ~.
\end{gather}
The above criteria should hold for all of the system's energy levels $i$. Note that the quantities written in Eq.~(\ref{criteria calc}) are exactly the terms leading to the ``slippage of initial conditions'', so finally we conclude that if the initial slippage for each eigenstate is small, then the perturbative treatment is a good approximation, otherwise one should look for different methods.

\section{Spin-boson response function}

In our simulation, the response function $\alpha(t) = D(t) + i D_1(t)$ is approximated by a series of exponents, obtained by a simple fitting procedure:
\begin{align}
\text{ps}^2 D(t) & \rightarrow \eta\times 
\\\notag& (0.14534\, +0.316206 i) e^{(-2.77201-0.985685 i) \hat t}
\\\notag&+(0.14534\, -0.316206 i) e^{(-2.77201+0.985685 i) \hat t}
\\\notag&-(0.0587924\, +0.0207246 i) e^{(-2.67694-3.11522 i) \hat t}
\\\notag&-(0.0587924\, -0.0207246 i) e^{(-2.67694+3.11522 i) \hat t},
\end{align}
\begin{align}
\text{ps}^2 D_1(t) & \rightarrow  \eta\times
\\\notag& -(0.00683011\, -0.0449112 i) e^{(-2.35315+1.04322 i) \tilde{t}}
\\\notag& +(0.00683011\, +0.00938383 i) e^{(-2.33632-3.21569 i) \tilde{t}} \\\notag& +(0.00683011\, -0.00938383 i) e^{(-2.33632+3.21569 i) \tilde{t}} 
\\\notag& +(-0.00683011-0.0449112 i) e^{(-2.35315-1.04322 i) \tilde{t}},
\end{align}
with $\tilde{t} = t/$ps. We find that for this case HEOM converges at $N_c=3$ for $\eta = 1$ and $N_c=9$ for $\eta = 10$.

\end{document}